\documentclass[twocolumn,preprintnumbers,amsmath,amssymb]{revtex4}

\usepackage{multirow}   
\usepackage{graphicx}
\usepackage{dcolumn}
\usepackage{bm}
\usepackage{epstopdf}
\usepackage{amsmath}

\begin{document}

\title{On the basic computational structure of gene regulatory networks}
\author{Carlos Rodriguez-Caso$^{1*}$ Bernat Corominas-Murtra$^1$ and  Ricard V. Sol\'e$^{1,2}$}

\begin{abstract}
Gene regulatory  networks constitute the  first layer of  the cellular
computation for cell adaptation and surveillance. In these webs, a set
of causal relations is built up from thousands of interactions between
transcription factors and their target genes.  The large size of these
webs and  their entangled  nature make difficult  to achieve  a global
view  of their  internal  organisation. Here,  this  problem has  been
addressed through a comparative study for {\em Escherichia coli}, {\em
  Bacillus   subtilis}  and   {\em   Saccharomyces  cerevisiae}   gene
regulatory networks. We extract  the minimal core of causal relations,
uncovering  the hierarchical  and  modular organisation  from a  novel
dynamical/causal  perspective.  Our results  reveal a  marked top-down
hierarchy  containing several small  dynamical modules  for \textit{E.
  coli}  and  \textit{B.   subtilis}.  Conversely, the  yeast  network
displays  a single  but  large dynamical  module  in the  middle of  a
bow-tie structure.  We found  that these dynamical modules capture the
relevant  wiring among  both common  and  organism-specific biological
functions such as  transcription initiation, metabolic control, signal
transduction,    response   to    stress,    sporulation   and    cell
cycle.   Functional   and  topological   results   suggest  that   two
fundamentally different  forms of logic organisation  may have evolved
in bacteria and yeast.
\end{abstract}

\affiliation{                    
  $^1$  ICREA-Complex Systems Lab, Universitat Pompeu Fabra (GRIB-PRBB). Dr Aiguader 88, 08003 Barcelona, Spain\\ 
    $^2$Santa  Fe Institute,  1399  Hyde Park  Road, New  Mexico
  87501, USA\\
  $*$ Corresponding author} 
  
\keywords{gene  regulatory networks,  modularity,  complex networks,  hierarchy,
Systems Biology}

\maketitle

\section{Introduction}

The pattern  of regulatory interactions  linking transcription factors
(TFs)  to  their  target  genes  constitutes  the  first  level  of  a
multilayered  network   of  gene   regulation;  the  so   called  gene
transcriptional regulatory  networks (GRN) \citep{Babu2004}.   Some of
these  patterns  have  been  recovered  from  genome-wide  approaches,
particularly   well    established   for   {\em    Escherichia   coli}
\citep{Thieffry1998,   Dobrin2004,Shen-Orr2002}   and  {\em   Bacillus
  Subtilis}   \citep{Harwood2002,Sellerio2009}    as   well   as   the
unicellular      eukaryote     {\em      Saccharomyces     cerevisiae}
\citep{Lee2002,Balaji2006}.   In  such  a picture,  both  hierarchical
\citep{Lagomarsino2007,Yu2006,Salgado2001}         and         modular
\citep{Resendis-Antonio2005,Wolf2003,Balaji2006,Freyre-Gonzalez2008}
components  have  been  repeatedly  highlighted, although  no  general
agreement exists  on what scale  of analysis more  accurately captures
global complexity \citep{Bornholdt2005}.

The  conceptualisation   of  cellular  interaction   maps  within  the
framework  of graph  theory  \citep{Albert2005,Zhu2007, Babu2004}  has
provided      powerful     insights     on      their     hierarchical
\citep{Clauset2008,Trusina2004,Vazquez2002,Sales-Pardo2007},    modular
organisation
\citep{Ravasz2002,Rodriguez-caso2005,Palla2005,Newman2004}  as well as
it sheds light on the topological constraints imposed in its dynamical
behaviour      \citep{Kauffman2003,Lee2002}.       However,      their
quantification,  even their  identification  has led  to a  nonuniform
concept  of  module under  functional,  topological, evolutionary  and
developmental  criteria  \citep{Wagner2007,Hartwell1999}.   Similarly,
the observed  hierarchy seldom matches an  ideal feed-forward relation
between components \citep{Whyte1969}.

An  alternative   approach  considers  looking  at  GRNs   in  a  more
fundamental way, namely  as sets of causal relations,  featured by the
directed nature  of the link  between a transcription factor  with its
target  genes \citep{Lagomarsino2005,Babu2004}.  Causal  links, namely
who acts  on whom, allows  to actually define the  skeleton underlying
dynamics by attending  to the cyclic and linear  nature of the genetic
circuits.
\begin{figure}
\begin{center}
\includegraphics[width=8cm]{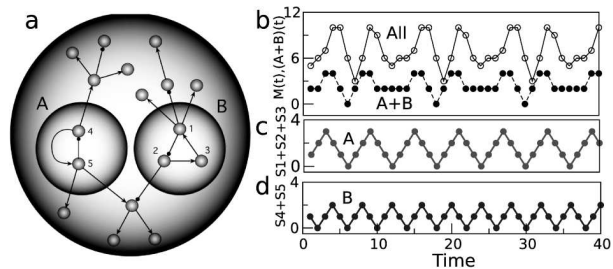} \label{diagrama}
\caption{A small toy model (a) and its dynamic behaviour. The observed
  pattern is  generated by the  activity of two basic  feedback loops,
  indicated as $A$  and $B$ in (a).  These  subsets are responsible for
  the qualitative dynamics exhibited by  the net, as shown in (b) with
  filled   circles.   Moreover,  this   attractor  results   from  the
  combination of  the two different  periodic orbits displayed  by the
  two basic modules, whose time series are shown in (c-d). For a mathematical 
  treatment of this specific example see appendix. }
\end{center}
\end{figure}

Figure \ref{diagrama}  illustrates the resulting dynamics of a toy model (see mathematical  appendix for formal
details).   In this model we observe that the  complex, periodic  behaviour  may arise
solely due  to the dynamics of the two  subgraphs (A and  B in
figure \ref{diagrama}a)  exhibiting feedback  loops.  The remaining system  simply reacts  to the  dynamical inputs  generated  from these
modules. Interestingly, these mentioned topological traits defines the dynamical behaviour no mattering what  kind of dynamical rules  are used to  causally relate our
elements.  Some  elements will play the leading  role, determining the
qualitative  type of  dynamics, whereas  others will  just  amplify or
reduce the core signals. This view of interactions is at the core of a
relational picture  of a biological  computation \citep{Rosen1991}. This
is a  computational perspective and, not surprisingly,  GRNs have been
compared           to           computers          \citep{Hogeweg2002,
  Istrail2007,Bray1995a,Rosen1991,Macia2009}.   Cellular  computations
pervade   both    the   diverse   responses    to   external   stimuli
\citep{Luscombe2004,  Balazsi2005}  as  well  as cell  robustness  and
plasticity  \citep{Aldana2007,Lee2007,Shmulevich2005}.   A  number  of
dynamical approximations  suggest a link  between network organization
and its dynamical  behaviour. However, due to the  large size of these
systems,  only a few  small ones  \citep{Li2004, Espinosa-Soto2004}
have been fully analysed.

Ideally, it would  be desirable to have a method  to construct a graph
capturing all  non-trivial causal  relations and thus  all potentially
important  computational  links.  In  this  paper  we  show that  such
organisation  of  GRNs  is  captured  by the  principle  of  causality
depicted by the directed nature of gene-gene relations. This can
uncovered  by  exploiting  the   properties  of  directed  graphs,  in
particular,   by   the  use   of   leaf   removal  algorithms   (LRAs)
\citep{Sellerio2009,Lagomarsino2005}  and  the  identification of  the
so-called strongly  connected components (SCCs)  \citep{Gross1998}. As
we will see,  LRAs recover the network fraction  responsible for potentially non
trivial  computations. Furthermore, the internal  organisation of  this especial
subgraph can be topologically  dissected by SCC identification.

A SCC
is defined for a set of vertices  if there is a \textit{walk} -attending to the
directedness of edges- from each  vertex to every other vertex of this
set (see figure \ref{SCC}). The identification of SCC has been applied
to  a diverse number  of systems  from the Internet  \citep{Broder2000} to
metabolism and neural wiring network \citep{Serrano2008}. Recently, it
has   been   described  that   yeast   GRN   presents   a  giant   SCC
\citep{Jeong2008a},  contrasting  with  an  acyclic  and  feed-forward
organisation of \textit{E. coli} \citep{Ma2004a}.

In  this work,  we explore  the topological  constraints  derived from
wiring patterns  of causal dependencies  in a comparative  analysis of
yeast, \textit{E.   coli} and \textit{B.  subtilis}  by defining their
qualitatively  relevant causal  cores.  We  explore  their internal
organisation  in terms  of irreducible  computational entities  at the
level of gene regulatory systems. As will be shown below, our analysis
(using  a more  updated  version of  \textit{E.   coli} GRN)  revealed
relevant  differences  in relation to   previous  results  published  in  the
literature.

\section{Results}


\subsection{Causality, dynamics and topology}

Causal relations in GRNs can  be described in terms of directed graphs
(in short \textit{digraphs}) \citep{Babu2004, Albert2005}.  A graph is
constituted by a  set of vertices or nodes (here the genes)  and the set of
edges linking them (the relations among genes).  The regulatory effect
of a TF gene $v_i$ on a specific target gene $v_j$  is depicted by an
arrow  ($v_i\rightarrow v_j$) in the graph plot. 

If the  vertices are  genes, a chain  of connections of  different TFs
corresponds in  graph theory  with a {\em  directed walk}\footnote{For
  the sake  of simplicity, we  present here the concepts  related with
  graph  theory  in an  intuitive  way.   For  formal definitions  and
  construction,  see  appendix.}   and  it  can be  interpreted  as  a causal chain. 
  Interestingly, all TFs exhibit outgoing
links, whereas non-TF genes (the target ones) only receive arrows from
the TF set.  The number of outgoing links of a vertex is known as {\em
  out-degree} (denoted  by $k_{out}$)  whereas the number  of incoming
edges  is  the  {\em in-degree}  ($k_{in}$).   Since  a  TF can  be  a
regulatory target  of other  TFs, they can  exhibit both  incoming and
outgoing links, allowing feed-backs to occur.

\begin{figure}
\begin{center}
\includegraphics[width=8cm]{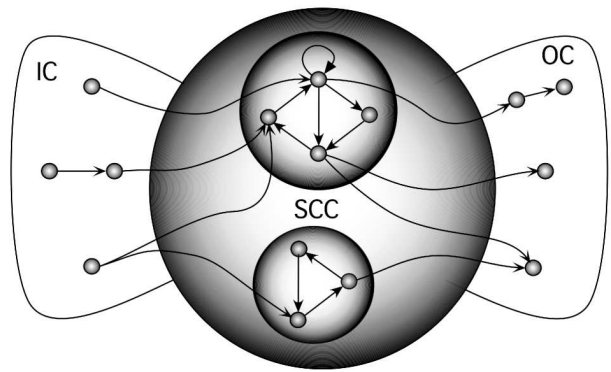} \label{SCC}
\caption{Bow-tie structure  depicting an incoming  component ($IC$), two
  strongly  connected  components (SCCs)  and  the outgoing  component
  ($OC$).}
\end{center}
\end{figure}

A specially important  information about the  organization of directed  graphs is
provided by the  analysis of its connected components,  i.e. the graph
in  which  every pair  of  distinct vertices  has  a  {\em walk}  -see
appendix.  In digraphs, directed walks are non trivially organised.   The Internet  is a  very illustrative
example of that \citep{Broder2000} exhibiting  a bow-tie structure with three different
connected  components:  1)  a  central  strongly  connected  component
(SCC), namely a  subgraph for which every two  vertices are mutually
reachable  \citep{Gross1998,Dorogovtsev2003},  2)  a set  of  incoming
components  ($IC$)  composed  by  the  set  of  feed-forward  pathways
starting from vertices  without in-degree and ending in  SCC and 3) a
set of  outgoing components ($OC$) where pathways  starting from SCC
end in vertices without out-degree (see figure \ref{SCC}).
Typically,  the  pattern  of  activation  of  a  given  gene  has  a
time-dependent  causal relation  with the  state of  the set  of genes
affecting it, and every vertex $v_i$  can acquire a given state from a
number of  possible states. Specifically,  the state of $v_i$  at time
$(t+1)$ is influenced by the  state of
the  vertices {\em  affecting}  it  at time  $t$  -see appendix.   The
equations describing the dynamical behaviour of a given vertex $v_i$ at
time  $t+1$ can  be formulated  in different  ways,  including Boolean
dynamics      \citep{Thomas1973,Kauffman1993},      threshold     nets
\citep{Li2004}  as   figure  \ref{diagrama}  illustrates   or  coupled
differential equations \citep{Fuente2002}.  These models are different
but all  show a common  principle of causality:  the state of  a given
vertex  $v_i$ at  time $t+1$  is exclusively  defined by  the vertices
affecting  it at  time $t$.   No matter  our choice  of  the dynamical
equations, the  patterning of  links must strongly  influence system's
behaviour.

The  presence of  a \textit{cyclic  walk} -a  directed walk  where the
first and  the ending vertices are  the same- is a  necessary (but not
sufficient) requisite for a periodic solution. This is due to the fact
that,  in interpreting directed  edges as  causal relations,  a cyclic
graph implies that every vertex is indirectly affected by itself. When
we consider the  whole graph, an overlapping of  cycles originates the
SCC set that can be governed by  the $IC$ set. As we shall show below,
these   two  structures  represent   the  core   of  the   graph  that
qualitatively  constrains its dynamical  complexity.  By  contrast, in
linear directed  walks, the  upstreamest element fully  determines the
final state of the elements belonging to this walk.

A set  able to properly capture the  relevant components qualitatively
affecting  global  behaviour  should  remove  linear  paths  from  the
graph.  Under this  view,  we can  identify  this set  by  means of  a
straightforward iterative algorithm. Given the  underlying influence  of causal  relations on
system's dynamics, we will use  the label \textit{dynamic} for most of
our definitions.

\subsection{Dynamical backbone}

\begin{figure}
\begin{center}
\includegraphics[width=8cm]{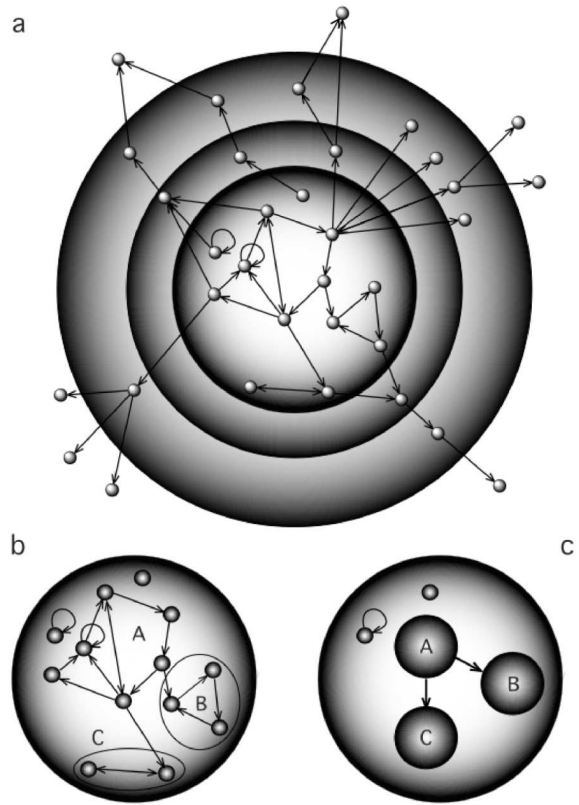}
\caption{Dynamical  Backbone  definition.    (a)  an  example  of  the
  iterative elimination  of vertices without  out-degree ($k_{out}=0$).
  Different layers indicate the successive iterations of the algorithm.
  In  (b) we  indicate the  dynamical modules  of the  $DB({\cal G})$.
  Maximal  cycles  larger than  one  vertex  are  labelled by  capital
  letters.   Figure (c)  shows  the hierarchical  organization of  the
  collapsed  dynamical modules,  i.e.,  the feed-forward  organization
  obtained after  the {\em condensation}  of the graph  $DB({\cal G})$
  -see  text.   Notice  that  the  $DB({\cal  G})$  exhibits  isolated
  self-interacting vertices that, by definition, are dynamical modules
  but auto-loops are maintained in order to provide a more informative
  picture.}
\label{DBB definition}
\end{center}
\end{figure}

We compute  the \textit{dynamical  backbone} ($DB$) of  a given  directed graph
${\cal G}$  (to be noted $DB({\cal  G})$) by the  iterative pruning of
vertices with $k_{out}=0$ from the initial graph, but maintaining those
vertices   whose   $k_{in}=0$   (figure  \ref{DBB definition}).    This
algorithm belongs to the  family the so-called leaf removal algorithms
\citep{Bauer2001,Correale2006,Lagomarsino2007}    that    have    been
previously  used   for  the  analysis   of  GRNs  \citep{Kauffman2003,
  Lagomarsino2007, Sellerio2009}.

Figure  \ref{DBB  definition}a illustrates  the  mechanism of  pruning.
Notice  that,  in contrast  with  previous  proposals of  leaf-removal
algorithms,  the  $DB({\cal  G})$  also  keeps the  {\em  single  root
  vertices} i.e.   those such  that $k_{in}=0$ which  appear isolated.
Single  root  vertices  are   special  because  their  state  is  only
externally changed and are not  influenced by other genes.  We observe
that,  in general  the  $DB({\cal  G})$ displays  more  than a  single
connected    component    (see    figure   (\ref{DBB    definition})).
Interestingly, this  dynamically relevant subgraph  coincides with the
union of  the previously described $IC$  and SCC sets  of a directed
graph.   By  contrast, $OC$  is  associated  with another  interesting
subgraph, to be indicated as $DB'({\cal G})$ formed by the fraction of
the net that exclusively displays feed-forward structures. See appendix 
for its formal construction.

The  application of  the above  defined algorithm  leads to  a drastic
reduction   of  network   complexity,  preserving   the  computational
structures responsible of the qualitative behaviour of the net.

\subsection{Dynamical Modules and Hierarchy} 

As we  argued above, the core  of causal relations of
 a GRN  is   captured   by  its   $DB$. Furthermore,  the causal  relations inside this  $DB$
can display some  kind of hierarchy. As we shall see $DB$ organisation  can be studied by SCC detention and the so-called graph \textit{condensation} process. Mathematical appendix formally details the treatment of GRNs that uncovers the  order relation (i.e. the hierarchical order) among the elements of the net.

Figure \ref{DBB definition}b illustrates the process for module identification by SCC detection. Interestingly, the nodes belonging to a SCC are mutually affected and therefore, they all contribute to the state definition of all the set  (see A and B regions of the toy model described in figure \ref{diagrama}a). The complexity of these topological entities cannot be reduced  and thus, SCC concept defines an irreducible  unit of  causal relations (see mathematical
appendix  for formal  definition). In  this sense,  the SCC can be identified 
with the  module concept, since it constitutes an identifiable semiautonomous
part of the system \citep{Wagner2007}, capturing a part of the dynamical
complexity of the  system. Therefore, we say that,  within the context of GRNs,  SCCs represent a sort of  dynamical modules.   We finally
note that, consistently with definition, a  single vertex belonging to the $DB$  but not  participating  in  any cycle  is  itself a  dynamical
module.

Interestingly, when  the SCC is represented by a single  vertex, the
resulting  graph is  a  directed acyclic  graph  (see figure  \ref{DBB
  definition}c).   Such  an  operation  is commonly  referred  in  the
literature as the {\em condensation} of a graph \citep{Gross1998}.

The   result  after   the  condensation   process  is   precisely  the
feed-forward organisation what enables  us to define an order relation
among the  elements of the $DB$. It is worth stressing that an order 
relation cannot be  established in cyclic wiring. Condensation process
 allows us to identify such cycles into SCCs representing dynamical 
 modules.  In this way, the internal organisation in a hierarchy of causal 
 relations is revelaled for the $DB$.

\subsection{{\em E. coli} dynamical backbone}
\begin{table*}[t]
\caption{A  selection of statistically over-represented biological  functions terms from gene ontology annotation (GOA) is shown for $DB$,
  the  root node  set  and  dynamical modules  (A  to E),  accordingly
  functional analysis for \textit{E. coli}. Notice that three genes of the $DB$ were not found in GOA annotation (see SI2 for a detailed analysis).  The number of genes belonging to a specific  GOA term is depicted in the first column. The fraction of genes associated to specific GOA respecting to the total number of genes for each particular set are shown for $P>0.05$  ($P$ is the level of significance). Highly significant results ($P<10E-5$) are indicated with (**).  Values within the interval of $10E-5>P>10E-3$  are marked by (*). }
\centering
\begin{tabular}{lcp{1.2cm}p{1.2cm}p{0.9cm}p{0.9cm}p{0.9cm}p{0.9cm}p{0.9cm}}
\hline
&&&&&&&&\\
Biological Function & $\#$ genes in GO &   \multicolumn{7}{c} {Fraction of the GO term in the analysed set}  \\ 
&&&&&&&&\\
 &  & \textit{E. coli}  DB
 & Root nodes & DM A & DM B & DM C & DM D & DM E \\ 
\hline
&&&&&&&&\\
Regulation of  & \multirow{3} {*} & & & & & & & \\ 
transcription, & 354 & 135/136* & 37/37* & 10/11* & 4/4** & 3/3 & 2/2 & 2/2  \\ 
DNA-dependent &&&&&&&& \\ 
&&&&&&&&\\
Regulation of metabolism & 399 & 136/136 & 37/37** & 11/11* & 4/4**
 & 3/3
 & 2/2 & 2/2 \\ &&&&&&&&\\
Two-component signal &  \multirow{3} {*} & & & & & & & \\ 
transduction system   & 91 & 26/136* & 6/37** & -- & -- & -- & -- & -- \\ 
(phosphorelay) & &&&&&&& \\&&&&&&&&\\
Transcription initiation & 7 & 7/136* & -- & 4/11* & 1/4 & -- & -- & -- \\&&&&&&&&\\ 
Negative regulation of &  \multirow{2} {*} & & & & & & & \\ 
cellular process & 36 & 10/136* & 4/37** & 2/11 & -- & -- & -- & -- \\&&&&&&&&\\ 
DNA replication & 62 & -- & -- & 3/11** & -- & -- & -- & -- \\ &&&&&&&&\\
Response to heat & 4 & -- & -- & 1/11 & -- & -- & -- & -- \\ &&&&&&&&\\
Response to antibiotic & 60 & -- & -- & -- & -- & 2/3 & -- & -- \\ &&&&&&&&\\
Carbohydrate transport & 112 & -- & -- & -- & -- & -- & 2/2 & -- \\&&&&&&&&\\
\hline
\label{Ecoli_table}
\end{tabular}
\end{table*}

We  extracted the  GRN for  {\em E.   coli} K-12  prokaryote  from the
available      information     in      RegulonDB      6.0     database
\citep{Gama-Castro2008}.  The  resulting network was  a directed graph
with 1607 vertices ($43$ of  them with $k_{in}=0$) and 4141 links with
a  giant component  of 1589  vertices  and average  degree $\langle  k
\rangle=5.1$  [See  Methods for  network  construction].  The  network
included a  total of 156 vertices with  $k_{out}>0$, corresponding with
transcription  and  $\sigma$   {\em  trans-acting}  factors,  and  1451
vertices with $k_{out}=0$, i.e, the target genes.

From  the {\em E.   coli} GRN,  the $DB$  subgraph was  obtained after
the second iteration of  the algorithm. The $DB$ subgraph  consists of $142$
vertices ($8.9\%$ of the GRN) distributed as follows: $33$ single root
vertices and  a set  of subgraphs with  $109$ vertices  ($10$ of
them with $k_{in}=0$)  and $279$ edges.  This set  is organised in $9$
graphs:  a  giant component  of  $100$  vertices  ($\langle k  \rangle
=3.7$),  another component  displaying two  elements and  $7$ isolated
self-interacting vertices. We found that  the size of the studied $DB$
is  markedly smaller  than the  expected in  a null  model  obtained by randomisation process. In this methodology, node
degree  is conserved  but correlations among nodes are destroyed (see methods section). Differences in the $DB$ size between \textit{E. coli} and randomised counterpart feature the biological fact of that only  a small
fraction of  genes in the genome operates over a majority of genes without a direct role in the regulation of transcription.
As  expected and in agreement with the direct observation of RegulonDB, the  genes belonging  to  $DB$ are  described as  either
transcription or  $\sigma$ factors.   Only two exceptions  were found:
the transcription  anti-terminator \textit{cspE} and  \textit{trmA}, a
tRNA metyltransferase (according to RegulonDB 6.0 [see SI for  a
 biological function  of $DB$  genes]). Functional analysis
based on gene ontology  annotation confirmed the significant ($P<10E-5$)
overabundance   of  general   functions  related   to   regulation  of
transcription, signal  transduction and transcription  initiation (see
table \ref{Ecoli_table} and SI2 for a detailed statistical analysis).

\begin{figure*}
\includegraphics[width=18cm]{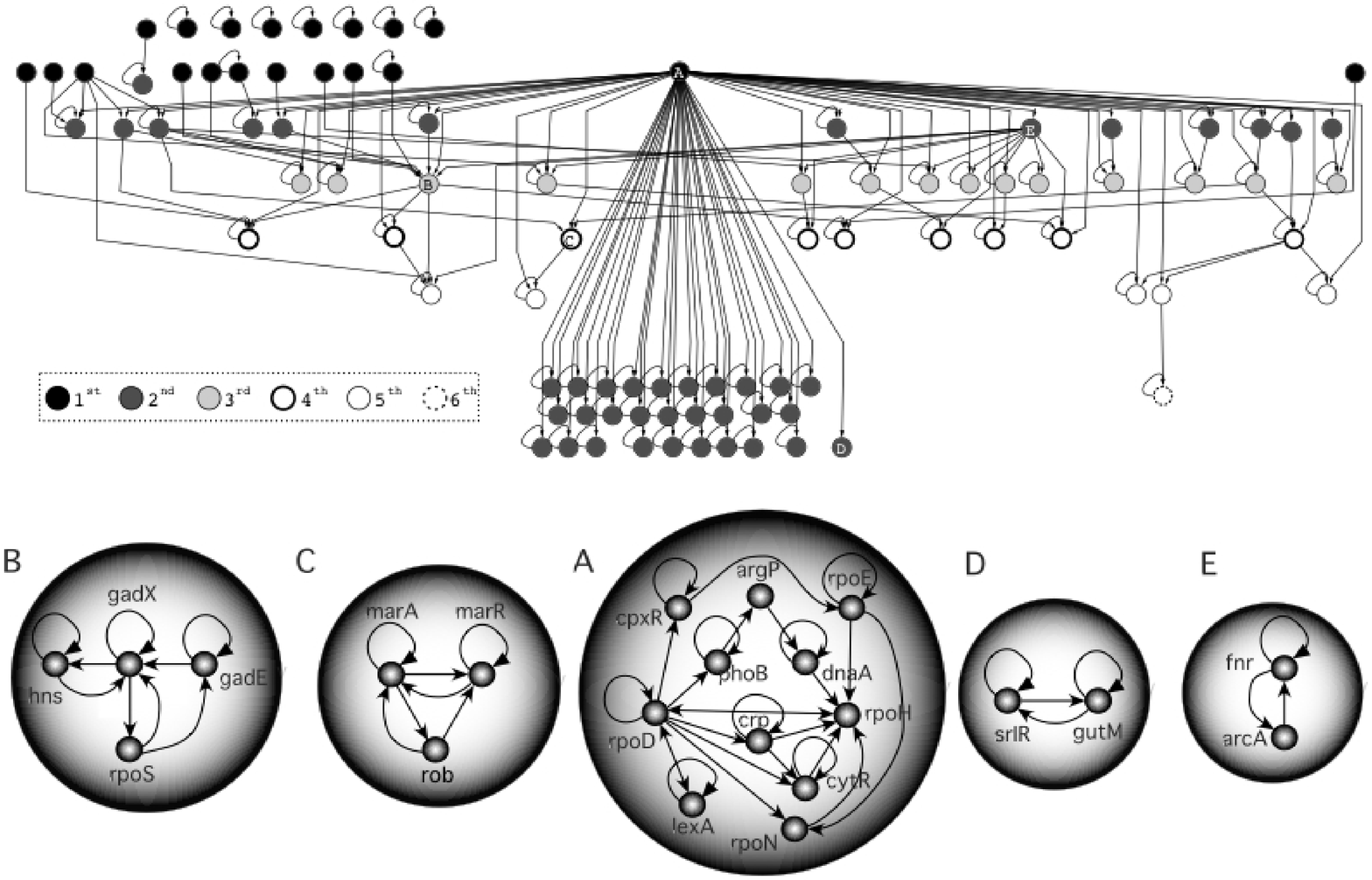}
\caption{Dynamical hierarchy  and modularity of  \textit{E. coli} GRN.
  Dynamical Backbone  of \textit{E.  coli}  after collapsing dynamical
  modules revealing  a causal  hierarchy (above). Different  levels of
  downstream  information   processing  are  indicated   by  means  of
  different vertex labelling. Dynamical modules larger than one vertex
  are represented  by an individual node labelled  by capital letters.
  Auto-links in  single-node module are  preserved in the graph  for a
  more informative picture.  Below {\em E.  coli} $DB$, the five (A to
  E) dynamical modules larger than one node are shown. }
\label{EcoliDBBdecomp}
\end{figure*}

\begin{figure*}
\includegraphics[width=18cm]{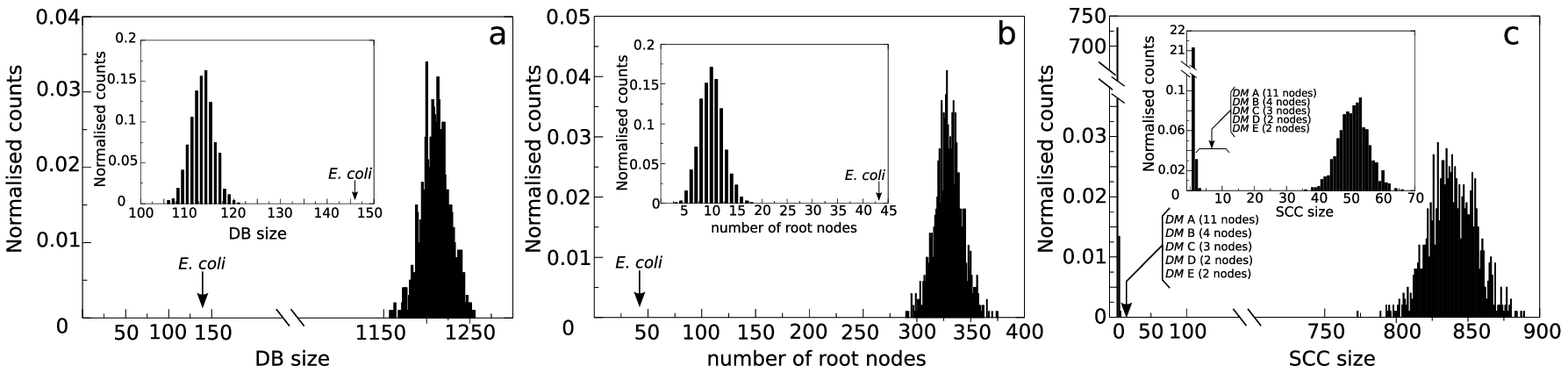}
\caption{Randomisation  for \textit{E.  coli} GRN  and  its respective
  TF-TF  subgraph (figure  insets). Normalised  counts  represent the
  number of  times that  a value of  the measured estimator  ($DB$ size,
  number  of root  nodes and  SCC size)  appears, normalised by  the  number of
  repetitions (1000 randomised networks)}
\label{Ecoli_rand}
\end{figure*}

Concerning the previous observations, we obtained a graph
 resulting from the interactions among TFs (topologically, the nodes with
  out-degree) in GRN. According its definition, $DB$ is also a subset of TF subnet. 
  Figure \ref{Ecoli_rand}a inset shows that $DB$ size for randomised TF-TF subgraph is fairly smaller than the  \textit{E. coli} $DB$. We also  found that the number of root  nodes expected by chance
in  both GRN  and  TF subgraph  follow  a similar  behaviour than  the
observed  for $DB$  size. These results show that $DB$
structure  for $E.  coli$ exhibits  a markedly  different organisation
than the expected in a null model (see figure \ref{Ecoli_rand}b).

Figure \ref{EcoliDBBdecomp}a shows the $DB$ organisation after the condensation process.  The
analysis of the condensed  $DB$ revealed five dynamical modules larger
than  one vertex  (figure  \ref{EcoliDBBdecomp}b). Interestingly,  the
size of SCCs are markedly smaller  in both GRN and TF randomised sets.
Notice that figure \ref{Ecoli_rand} depicts large values for  SCC sizes equal to one in randomised networks, however, they correspond with the fraction of the non condensed network. Typically, randomised networks exhibited a single large 
SCC component (within the range of 800-900 nodes), occasionally together  with a very small SCC, rarely larger than three nodes (data from SCC distributions within networks are not shown). It is worth to note that  no  randomised web,  neither  for GRN  nor   TF-TF subgraph  sets produced exclusively small SCCs as observed in real data. This  indicates that  GRN is  not devoid of  cycles but  they are
distributed along the $DB$ and confined in dynamical modules of small size.

We found  that the  \textit{condensed} $DB$ captures  the hierarchical
behaviour of  the largest graph component evidenced  by a feed-forward
order  relation  with  six  layers  of  downstream  dependencies (see figure \ref{EcoliDBBdecomp}).   By
definition, the first layer  contains all the vertices with $k_{in}=0$
but we can see that it also includes the largest dynamical module.

The largest module (A in figure \ref{EcoliDBBdecomp}) contains four of
the seven  $\sigma$ factors  described for {\em  E.  coli}  (see table
\ref{Ecoli_table} and SI2 for statistical details).  These elements  are responsible for transcription
initiation.  Together with  the primary initiator factor \textit{rpoD}
($\sigma^{70}$), we  found the  alternative ones operating  under heat
shock  stress  (\textit{rpoH}  and \textit{rpoE},  corresponding  with
$\sigma^{32}$   and   $\sigma^{24}$,   respectively).   In   addition,
\textit{rpoN} ($\sigma^{54}$, initiator  of nitrogen metabolism genes)
is also part of the module.  The second largest \textit{hub} (a highly connected vertex) in the $DB$ is the
\textit{crp} gene,  also known as CAP  (catabolite activator protein).
CAP is a  general regulator that exerts a positive  control of many of
the   catabolite   sensitive   operons   as  a   sensor   of   glucose
starvation.  Interestingly,  the
members of  module A participates in the  $44\%$ of the  total number of links of the GRN, indicating
the  relevance of  this  module in  defining  the state  of the  whole
network.

Other relevant  factors are co-localised  in this dynamical  module
such  as  those  TFs  related  to  nutrient  sensor  and  assimilation
(phosphate  sensor  system,   \textit{phoB},  as  well  as  nucleoside
(\textit{cytR}) and arginine (\textit{argP}) transport control.  Notice  that the  initiator  factor of  DNA replication  initiator
(\textit{dnaA}) is associated  with this group, and we  also found two
specific  TF  expressed  under  stress conditions  (\textit{lexA}  and
\textit{cpxR}).  Similarly,  the other four modules  include key genes
associated  to  adaptive responses  to  changing environmental  clues.
These  include homeostasis  in acid  environment (stress  responses to
high   osmolarity,   module  B),   antibiotic   resistance  (lead   by
\textit{marA} and \textit{marR}, module C), glucitol use (module D) or
responses to oxygen changes (module E).

In  summary,  the  \textit{E.   coli} $DB$  describes  a  hierarchical
feed-forward  network with a  strong fan-like  pattern dominated  by a
single irreducible subset of TFs.  This pattern reveals, in terms of a
computational design, a strongly centralised organisation, with a well
defined node  affecting multiple layers  of activity. The  largest hub
has no  input from other members of  the $DB$ and thus  only affect to
other's behaviour.  Within the $DB$, we identify the dynamical modules
responsible for the control  of transcriptional replication under both
normal and  stress conditions, control of  metabolism, DNA replication
as  well  as  assimilation   of  essential  sources  of  nitrogen  and
phosphorous.

\subsection{{\em B. subtilis} dynamical backbone}

\begin{table*}
\caption{A  selection of statistically over-represented biological  functions terms from gene ontology annotation (GOA) is shown for $DB$,
  the  root node  set  and  dynamical modules  (A  to E),  accordingly
  functional analysis for \textit{B. subtilis}. The number of genes belonging to a specific  GOA term is depicted in the first column. Notice that 8 genes of 144 belonging to $DB$ were not found in GOA and therefore not considered in this analysis. The fraction of genes associated to specific GOA respecting to the total number of genes for each particular set are shown for $P>0.05$  ($P$ is the level of significance). Highly significant results ($P<10E-5$) are indicated with (**).  Values within the interval of $10E-5>P>10E-3$  are marked by (*). }
\centering
\begin{tabular}{lcp{1.2cm}p{1.0cm}p{0.9cm}p{0.9cm}p{0.9cm}p{0.9cm}p{0.9cm}}
\hline
&&&&&&&&\\
Biological Function & $\#$ genes in GO &   \multicolumn{7}{c} {Fraction of the GO term in the analysed set}  \\ 
&&&&&&&&\\
 &  & \textit{B. subtilis}  DB
 & Root nodes & DM A & DM B & DM C & DM D & DM E \\ 
\hline
&&&&&&&&\\
Regulation of  & \multirow{3} {*} & & & & & & & \\ 
transcription, & 357 & 113/121* & 58/62* & 2/2 & 2/2 & 2/2 & 2/2 & 3/3 \\
DNA-dependent  &&&&&&&&\\
&&&&&&&&\\
Regulation of metabolism
 & 368 & 114/121* & 58/62* & 2/2 & 2/2 & 2/2 & 2/2 & 3/3 \\ 
&&&&&&&&\\
Two-component signal & \multirow{3} {*} & & & & & & & \\ 
transduction system & 99 & 8/121* & 12/62* & -- & -- & 2/2 & -- & -- \\ 
(phosphorelay)&&&&&&&&\\
&&&&&&&&\\
Transcription initiation & 26 & 14*
 & 5/62** & 2/2 & 2/2 & -- & -- & 1/3 \\ 
&&&&&&&&\\
Negative regulation of &  \multirow{2} {*} & & & & & & & \\ 
cellular process & 15 & 8/121* & 3/62 & -- & -- & -- & -- & -- \\ 
&&&&&&&&\\
Positive regulation of &  \multirow{2} {*} & & & & & & & \\ 
cellular process & 8 & 6/121* & 6/62* & -- & -- & -- & -- & -- \\ 
&&&&&&&&\\
Transcription termination & 6 & 3/121 & 2/62** & -- & -- & -- & -- & -- \\ 
&&&&&&&&\\
Sporulation & 
258
 & -- & -- & -- & 2/2
 & -- & -- & 3/3
 \\ 
&&&&&&&&\\
Phosphate transport & 10 & -- & -- & -- & -- & 1/2 & -- & -- \\ 
&&&&&&&&\\
\hline
\label{Bsubtilis_table}
\end{tabular} 
\end{table*}

\textit{B.  subtilis}  GNR was obtained  from DBTBS \citep{Sierro2008}
       [see Methods for  network construction].  The resultant network
       consists of  922 nodes (159  with $k_{out}>0$, 83 of  them with
       $k_{in}=0$ and  763 with  $k_{out}=0$) and 1366  directed links
       ($\langle  k\rangle =  2.96$).  A  component of  892  nodes was
       predominant and  the remaining nodes  made part of  very little
       isolated subgraphs (one of five  nodes, one of three, 10 of two
       nodes and two isolated self-interacting nodes).

The  computation of  $DB$ required  four iterations  before the acquisition of the stable graph. The  process removed 778  nodes, resulting a $DB$  of 144
genes (10.5 \% of the GRN) .  In this set, 73 were isolated root nodes
and 71 nodes were distributed in  8 connected components (a big one of
59  nodes, one of  five, one  of two  and 5  self-interacting isolated
nodes) with 108 links. Analogously to the observed for \textit{E. coli}, statistical analysis of GOA terms showed and over-abundance of regulation of transcription metabolism and cell signalling (see table \ref{Bsubtilis_table} and SI2 for statistical details) in both $DB$ and root nodes set (see table \ref{Bsubtilis_table} and SI2 for statistical details). Other functions relating to transcription termination and positive regulation process appear in \textit{B. subtilis} $DB$.

We note that comparison of functional analysis among species must be taken with caution due to the different coverage of genes in GOA terms that may produce a bias in the interpretation of results.  In order to provide a more complete picture, we consider functional information annotated in DBTBS. However, we cannot exclude some discrepancies derived from the use of different sources.  Similarly to \textit{E. coli}, statistical functional analysis is complemented by a direct  function examination of the genes belonging to $DB$ in DBTBS (see SI for functional information).

\begin{figure*}
\includegraphics[width=18cm]{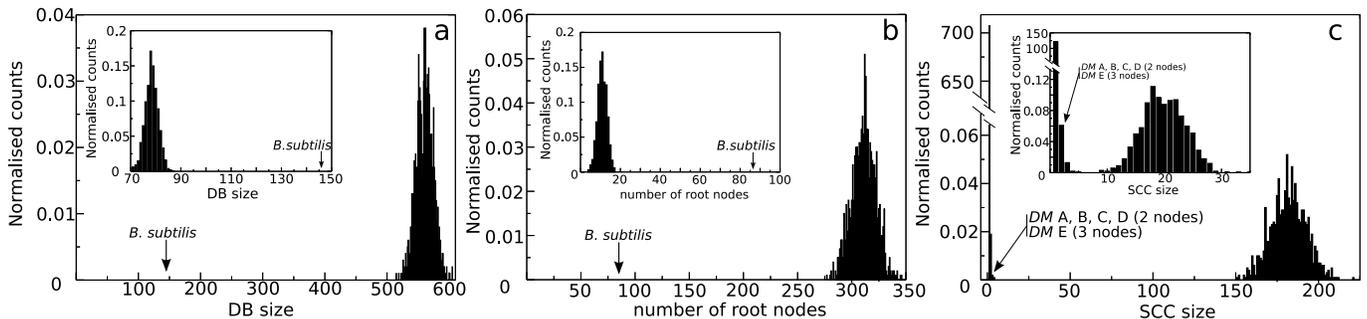}
\caption{Randomisation for \textit{B. subtilis} GRN and its respective
  TF-TF  subgraph (figure  insets). Normalised  counts  represent the
  number of  times that  a value of  the measured estimator  ($DB$ size,
  number  of root  nodes and  SCC size) appears normalised by  the  number of
  repetition (1000 randomised networks)}
\label{B.subtilis_rand}
\end{figure*}

Concerning to null model comparison, figures  \ref{B.subtilis_rand}a and b suggest a similar behaviour for $DB$ and the set of root nodes in randomised networks than the observed for \textit{E. coli}.

\begin{figure*}
\begin{center}
\includegraphics[width=18cm]{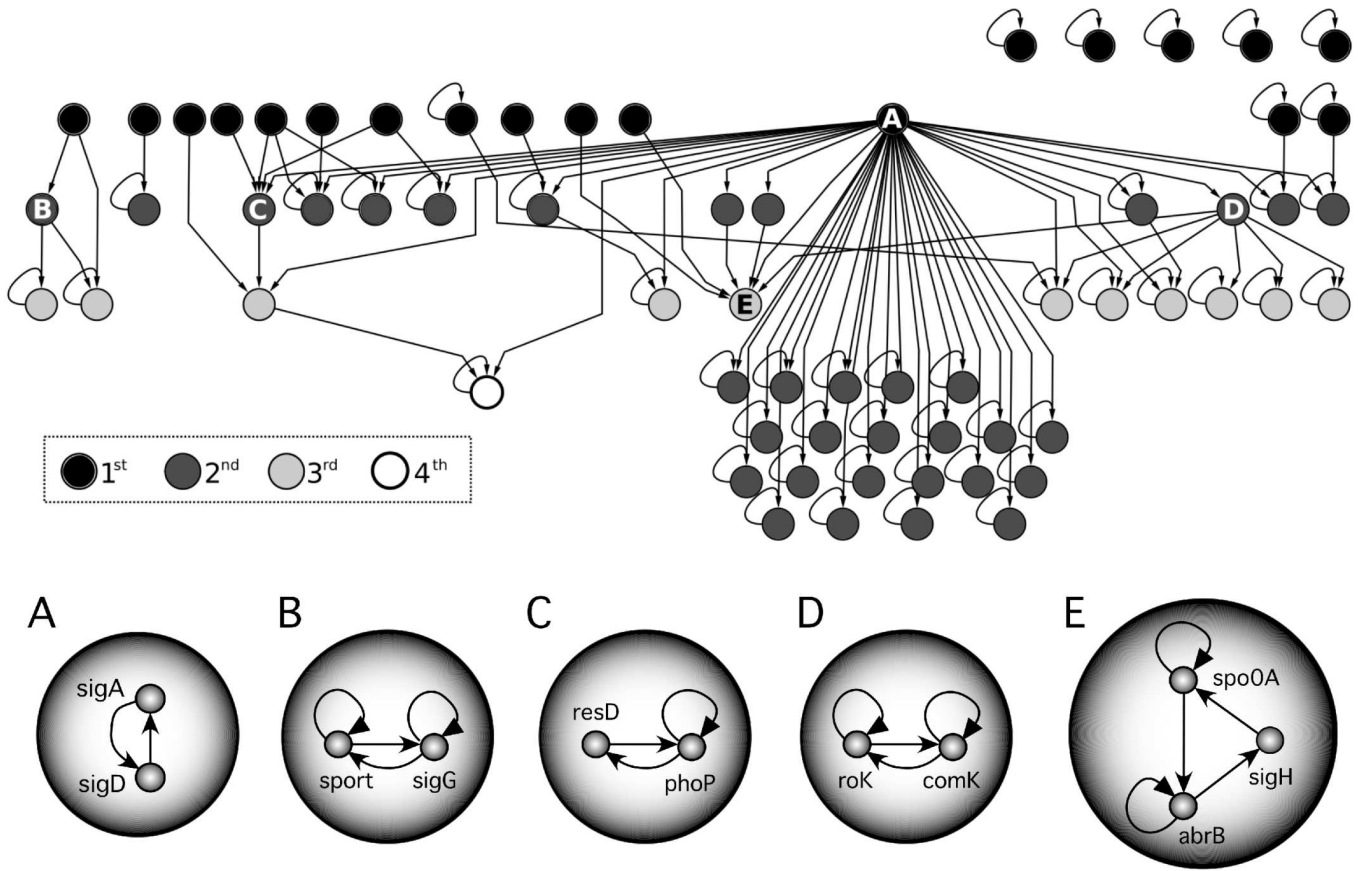}
\caption{Dynamical  hierarchy and  modularity of  \textit{B. subtilis}
  GRN.   Dynamical Backbone of  \textit{B. subtilis}  after collapsing
  dynamical  modules revealing a  causal hierarchy  (above). Different
  levels of  downstream information processing are  indicated by means
  of  different vertex  labelling. Dynamical  modules larger  than one
  vertex are  represented by an individual vertex  labelled by capital
  letters.   Auto-links in  single-node  module are  preserved in  the
  graph for a more informative picture.  Below {\em B. subtilis} $DB$,
  the  five  (A to  E)  dynamical modules  larger  than  one node  are
  shown. }
\label{figuraSCCsubtilis}
\end{center}
\end{figure*}

SCC  calculation revealed  five  dynamical modules  as illustrated  in
figure \ref{figuraSCCsubtilis}. Four of  them in the same 54-connected
component and the other one in the 5-component component. Similarly to \textit{E. coli}, a single markedly large SCC is expected in randomised networks (see figure \ref{B.subtilis_rand}c). In the same way, the acyclic
graph resulting by SCCs condensation revealed a strong
hierarchical  character, similar  to its  \textit{E.   coli} bacterial
counterpart.  In this case, module A consisted of a cross-regulation of
\textit{sigA}   and  \textit{sigD}   sigma   factors. These two nodes participates in the $26\%$ of the total number of links of the GRN.  This agrees with the essential role of these genes.
\textit{SigA},  also known  as $\sigma^{43}$  or \textit{rpoD},  is an
essential gene, primary $\sigma$ factor of this bacterium.  It is also
worth  nothing  that  the  equivalent \textit{rpoD}  gene  occupies  a
similar  position on  the top  of the  \textit{E.  coli}  condensed $DB$
(module  A in  figure \ref{EcoliDBBdecomp}).   The module
partner  of  this  gene  in \textit{B.   subtilis},  \textit{sigD}  or
$\sigma^{28}$, is  a sigma factor required for  flagellum and motility
genes  involved  in  chemotaxis   process.   This  is  different  from
\textit{E.  coli},  since  its  equivalent  gene  \textit{fliA}  (also
$\sigma^{28}$) does not  belongs to the \textit{E. coli}  module A but
it has a direct downstream dependence of this module.

As  illustrated in  figure  \ref{figuraSCCsubtilis}, modules  C and  D
receive a direct  input from the master module  A. They appear related
to  signal transduction  systems (\textit{resD}  and  \textit{comk} in
module  C and  D,  respectively).  Interestingly,  a cross  dependence
between respiration  and phosphate assimilation is  captured in module
C. It has been described that \textit{resD} can also play a regulatory
role in  respiration whereas \textit{phoD}  makes part of  a molecular
system  involved  in  the   regulation  of  alkaline  phosphatase  and
phosphodiesterase  alkaline phosphate,  two enzymes  involved  in the
uptake  of free phosphate  groups from  the environment.   Moreover, a
catabolite  repression  factor,  \textit{ccpA},  responsible  for  the
repression of carbohydrate utilisation and the activation of excretion
of excess carbon affects this module.  This suggests a strong relation
between   phosphorous  and   carbon   assimilation  with   respiratory
regulation.   In  other hand,  the  \textit{rok-comk} genetic  circuit
described in module  E appears involved in the  expression of the late
competence genes.

In  contrast to  what occurs  in  \textit{E. coli},  the most  complex
module  is not  at the  top of  the hierarchy.   Module E  is  a three
component  gene  (much smaller  than  for  \textit{E.  coli})  closely
related  to sporulation process.   It is  noteworthy that  this module
receives  inputs  from the  competence-related  module  E and  another
sporulation module (see table \ref{Bsubtilis_table}) is found in an independent component. This is the case
of module B which does not belong to the same $DB$ connected component
of the remaining modules.  It contains a sporulation-specific $\sigma$
factor   (\textit{sigG},  together  with   \textit{spovT}  controlling
factor).    Interestingly,  this   module, significantly associated to sporulation function (see Table \ref{Bsubtilis_table}),  is   affected   by  another
sporulating $\sigma$  factor (\textit{sigF}).  Another  relevant trait
is  the link between  sporulation and  response to  nutritional stress
performed  by \textit{spo0A}  and  the catabolite  repression role  of
\textit{abrB} in module D.

From a topological point of  view, the most obvious difference between
\textit{E.  coli} and \textit{B.  subtilis} is the simpler composition
of dynamical modules  for bacillus. We must stress  that the knowledge
of these organisms -specially for \textit{B. subtilis}, which is
much  less  known  that  its  E.  coli  counterpart-  is  in  constant
advance. Thus,  a variation in the  wiring of these  modules cannot be
neglected.  However, differences in  module conformation might also be
due to the  large phylogenetic distance among these  bacteria, as well
as   the   extremely   different   environments   where   they   live.
\textit{E. coli} is a gram-negative enterobacteria unable to sporulate
that   lives    in   the   intestine   of    warm   animals,   whereas
\textit{B. subtilis} is a  free living gram-positive organism, usually
found in soils  and with the ability to  sporulate.  However, in spite
of  these  differences, a  very  similar  architecture  with a  strong
hierarchical character  of its $DB$  is shared.  Once again,  a single
only-output hub is observed. As show below, this is markedly different
from what is found in our eukaryotic example.

\subsection{Yeast dynamical backbone}

\begin{figure}
\includegraphics[width=8cm]{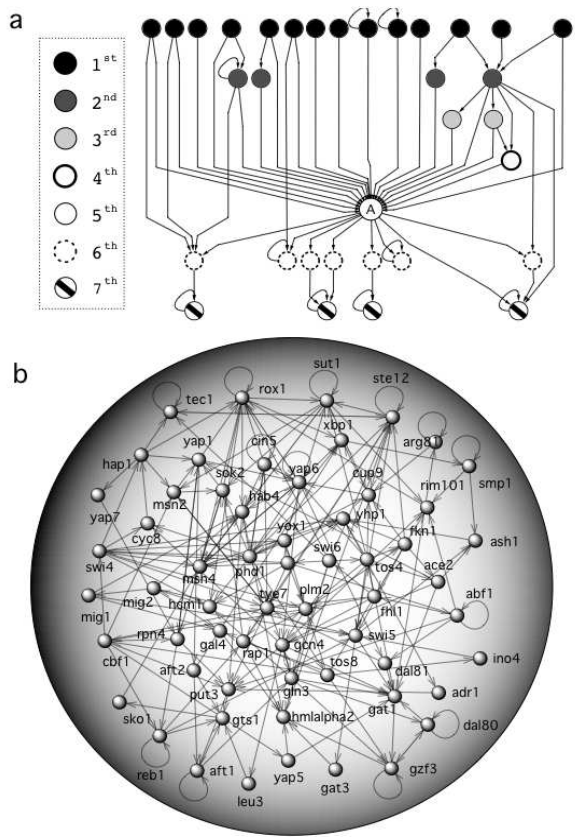}
\vspace{0.25 cm}
\caption{Dynamical  hierarchy and modularity  of \textit{Saccharomyces
    cerevisiae}  GRN.  Dynamical  Backbone of  yeast  after collapsing
  dynamical   modules  (above).   One   single  dynamical   module  is
  represented  by  a  vertex  labelled  by A  letter.   Below,  the  A
  dynamical module.  As in  the other diagrams,  box legend  shows the
  levels of computation.}
\label{yeastDBBdecomp}
\end{figure}

\begin{figure}
\begin{center}
\includegraphics[width=6cm]{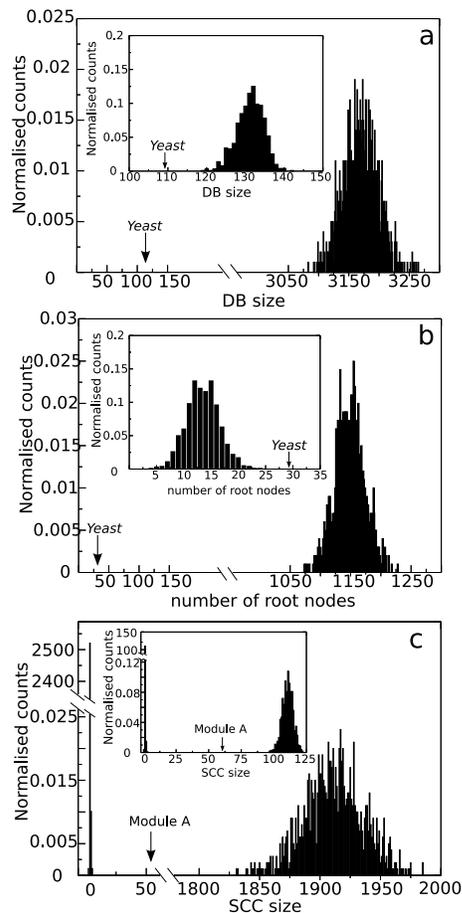}
\caption{Randomisation for \textit{B. subtilis} GRN and its respective
  TF-TF  subgraph (figure  insets). Normalised  counts  represent the
  number of  times that  a value of  the measured estimator  ($DB$ size,
  number  of root  nodes and  SCC size) appears normalised by  the  number of
  repetition (1000 randomised networks)}
\label{yeast_rand}
\end{center}
\end{figure}

GRN of  {\em Saccharomyces cerevisiae}  eukaryote was obtained  from a
compilation  of   different  genome  scale   transcriptional  analyses
\citep{Balaji2006}  [see  Methods   for  network  construction].   The
resulting  network  was  a  directed  graph  consisting  of  a  single
connected component  with 4441 vertices  (29 of them  with $k_{in}=0$)
and  12900 links,  leading to  a $\langle k  \rangle=5.8$.   The network
included a total of 157  vertices with $k_{out}>0$ and $4284$ vertices
with  $k_{out}=0$, corresponding  with the  transcription  factors and
targets  genes   analysed  in  the  different   datasets  compiled  in
\citep{Balaji2006}.

The yeast  $DB$ subgraph was  obtained after three iterations  of $DB$
pruning. The $DB$ was formed by a set of $109$ vertices
($2.45\%$  of  the GRN),  being  $17$  of  them isolated  single  root
vertices, and a single  connected component displaying 92 vertices and
318 edges, leading to $\langle k \rangle =6.1$. Analogously to bacteria GRNs, randomisation process  produced networks with much larger $DB$ with a smaller number of root nodes and a very large SCC compared to real yeast $GRN$ (see figure \ref{yeast_rand}). However, a different behaviour was observed from yeast TF-TF subgraph producing a fairly bigger $DB$ after randomisation process. This suggests that although bacteria
and yeast $DB$  exhibited a similar size, they  would differ in their $DB$
organisation. 

Functional analysis revealed that $DB$ consists of elements responsible for regulation of transcription and metabolism as observed in bacteria. In addition cell cycle related functions and response to chemical stimulus appeared significantly over-represented in  $DB$, the root node set and SCC (see table \ref{yeast_table} and SI2 for detailed statistical analysis).

All the  109 genes of the  $DB$ appear included in  the list obtained
from \citep{Balaji2006}  defining the TF  genes of the yeast  GRN.  In
addition,  we  checked  that  all  $DB$  elements,
but five were clearly identified  as TFs in the current version
Saccharomyces  genome database \citep{Hong2008}.   However,
they all were classified as  regulators of gene expression [See SI for
  biological details of $DB$ genes].

The condensation process  of yeast $DB$ revealed a single
dynamical   module   of   $60$    vertices   (module   A   in   figure
\ref{yeastDBBdecomp}a) with a high  average degree ($\langle k \rangle
= 6.6$) that contrasts to SCC distribution in bacteria.  The  module actually included more than a  half of the $DB$,
tied    to    a   much    less    marked    hierarchy   (see    figure
\ref{yeastDBBdecomp}b)  than  observed in  the  {\em  E.  coli}  $DB$.
However,  the  number  of  layers  (seven)  was  very  close  in  both
organisms, as figure \ref{yeastDBBdecomp}c indicates.

We can see that the large dynamical module occupies a central position
within  the $DB$.   Overall, the  yeast $DB$  resembles  the so-called
bow-tie organisation  observed in the  Internet \citep{Broder2000} and
metabolism \citep{Ma2007}  where incoming  inputs are integrated  in a
large  component   leading  to  a  set  of   outgoing  outputs.   This
substantially differs from the  hierarchical character of {\em E. coli
  DB} and  from the  point of  view of computation  it reveals  a much
higher integration and pre-processing of the input-output paths.

Interestingly,  module A  contains  a marked over-abundance of cell cycle functional terms ($P<10E-5$, see table \ref{yeast_table} and SI2 for detailed statistical analysis). Direct observation of database information indicated that  TFs for  cell  cycle such  as
\textit{swi4},  \textit{swi6} and  \textit{swi5}  (they control  $G_1$
related  genes   and  they  are   also  involved  with   DNA  repair),
\textit{hcm1}   ($S$   phase   related   genes),   \textit{yhp1}   and
\textit{yox1} ($M/G_1$ phase) and \textit{fkh1-2} ($G_2$ phase), among
others.  Moreover, the swi TFs located  in module A make part of the 5
TFs  contained in the  minimal dynamical  network suggested  for yeast
cell  cycle \citep{Li2004}.   Although this  network  combines protein
modification  besides transcription, we  observe the  all TFs  in Li's
network are included in the yeast  $DB$. The two remaining TFs in Li's
module  that  are  not  inside  module  A  (\textit{mcm1,  mbp1})  are
upstreamly located in the  hierarchy, indicating a master control over
the   swi  factors   (and  the   remainder  module   A   partners)  at
transcriptional level.   It is also  found a number of  TF controlling
the   assimilation  of   carbon  sources,   amino   acid  assimilation
(\textit{gal,  adr1,  aye7,  myg1-2,  put3,  arg81,  leu3,  gln3}  and
\textit{gcn4}), nitrogen compound degradation (\textit{dal80-81, gzf3,
  gat1}), stress  response (\textit{msn4,sut1,yap1-6, sko1,  smp1} and
\textit{msn2})   and   sporulation   (\textit{rim101}).    A   detailed
description  of biological  functions derived form manual analysis of  module A  and yeast  $DB$ are
detailed in SI.

\begin{table*}[t]
\caption{A  selection of statistically over-represented biological  functions terms from gene ontology annotation (GOA) is shown for $DB$,
  the  root node  set  and  dynamical modules  (A  to E),  accordingly
  functional analysis for \textit{S. cerevisiae}. Notice that only two genes of the $DB$ were not found in GOA annotation (see SI2 for a detailed analysis).  The number of genes belonging to a specific  GOA term is depicted in the first column. The fraction of genes associated to specific GOA respecting to the total number of genes for each particular set are shown for $P>0.05$  ($P$ is the level of significance). Highly significant results ($P<10E-5$) are indicated with (**).  Values within the interval of $10E-5>P>10E-3$  are marked by (*). }
\begin{tabular}{lcp{1.54cm}cc}
\hline
&&&&\\
Biological Function & $\#$ genes in GO &  & Fraction of the GO term &  \\ 
 &  & \textit{S. cerevisie} $DB$ & Root nodes & DM A \\ 
\hline
&&&&\\
Regulation of metabolism & 466
 & 80/107* & 14/27* & 51/60* \\ 
&&&&\\
Regulation of transcription, DNA-dependent & 327 & 69/107* & 10/27**
 & 44/60* \\ 
&&&&\\
Positive regulation of cellular process & 101 & 33/107* & 7/27* & 21/60* \\ 
&&&&\\
Negative regulation of cellular process & 191 & 17/107* & --
 & 13/60* \\ 
&&&&\\
Response to chemical stimulus & 270 & 20/107* & 5/27 & 11/60** \\ 
&&&&\\
Cell cycle & 412 & 21/107**
 & -- & 17/60* \\ 
&&&&\\
Transcription initiation & 56 & 5/107 & -- & -- \\ 
&&&&\\
G1/S-specific transcription in mitotic cell cycle & 12
 & 9/107* & -- & 9/60* \\ 
&&&&\\
\hline
\label{yeast_table}
\end{tabular} 
\end{table*}

\section{Discussion}

The final goal  behind any network-based analysis is  in most cases to
reach a systematic picture of {\em  what is relevant} and what is not.
According to this philosophy, we have presented a comparative study of
the best  known up-to-date GRNs  by exploiting their  most fundamental
trait: the principle of causality. The GRN can be seen as a roadmap of
the  computational processes  linking external  stimuli  with adaptive
cell responses and this can be reduced to the wiring pattern of causal
dependencies  among TFs and  their target  genes.  Although  we ignore here any 
dynamical features, stability properties and other relevant components
of GRN  complexity, the construction process is  unambiguous and leads
to a unique, logical description of the causal core of a GRN.

By using  the directed  nature of the  network, we  can systematically
reduce network complexity from thousands of elements to a much smaller
subset including  all the causally relevant  parts of the  GRN. Such a
causal/computational  perspective  is consistent  with  the nature  of
regulatory maps. No matter how  they behave exactly (and thus how they
would  be  modelled) the  set  of  non-reducible  causal links  should
capture  most of  what is  dynamically relevant.  The approach
  presented   here  combines   the   use  of   leaf-removal  and   SCC
  identification  algorithms for directed  networks. These  methods do
  not  make  any assumption  or  require  any probabilistic  estimator
  providing  a  unique solution where  the  presence  of  a  modular
  hierarchy can be easily depicted.

This  approximation to the identification of the basic computational/causal units differs from  others  such as
  network  motif  identification.  Dissecting  networks  into  motifs provides  a  systematic
  classification   of   GRNs    in   their   small   building   blocks
  \citep{Shen-Orr2002}. By using the abundance of small subgraphs in a given network. However, the detection of a subgraph in a given net requires predefined decisions on what is to be measured.  By  contrast,  SCC detection is not restricted to the search of a particular subgraph configuration and makes possible the identification of entities that cannot be further partitioned due to the cyclic behaviour of this special set.  As we stated in the introduction section, cycles are
  at the  basis on  a non-trivial computation  \citep{Thieffry2007} and, as we have shown, SCC is a topological concept that
  may  be  interpreted as  a  sort  dynamical modules  (identifiable
  semiautonomous entities \citep{Wagner2007}) within the framework of GNR. Attending to this, we have  shown that these dynamical modules are organised  within the core of  computationally relevant dependencies that is the
  $DB$. Interestingly, this interpretation is also valid for any network depicting a relation of causal dependencies.

The analysis  of GRNs  shows that dynamical  backbones are  very small
(compared with the whole network) and thus the basic logic of GRNs can
be strongly  simplified. The very small number  of iterations required
to obtain  the $DB$ suggests a direct  control of the TF  set over the
pool  of  target  genes,  without  a long  \textit{chain  of  command}
responsible for  information transmission.  This is  reinforced by the
fact that in  all cases more than the $90\%$  of genes were eliminated
after the first iteration of $DB$ pruning (data not shown).
Functional analyses did not reveal significant differences between $DB$ and the set of root nodes. Since they are at the top of hierarchy, we could conclude that they might be related to stimulus sensing. However, we must stress that the TF-gene target relation is a part of the regulation process and TF and promoter alteration through cell signalling processes or metabolism can operate in different elements of the regulatory set. In this context, further efforts in the integration of different regulatory levels would contribute to clarify this issue. 

The  strong  hierarchical  $DB$  structure found for  bacteria  is consistent  with
previous    observations   \citep{Sellerio2009,Ma2004a,Yu2006}.    Our
analysis   reveals  that   bacterial   GRNs  are   almost  devoid   of
cycles, although not totally excluded,  as previous
studies have suggested  \citep{Ma2004a}. Moreover, our  results indicate that
these  cycles  are organised  into  dynamical  modules with  essential
biological  functions, such  as module  A in  \textit{E.   coli}.  
The
pattern  of   connections  of  this  module  reveals   a  non  trivial
computational  link  between transcription  and  metabolic control  under both
normal and stress growth conditions. As  we have shown, it is a common
shared trait  in bacteria despite the relative  simple organisation of
\textit{B.   subtilis}  when  compared  with \textit{E.   coli}  (most
probably  due to  a lesser  knowledge of  its regulatory  wiring).  In
contrast  with  the  prokaryotic  $DB$s,  the  yeast  $DB$  is  fairly
different, with a  high content of cycles organised  into a single SCC,
(see \citep{Jeong2008a}). Our analysis shows that this large dynamical
module  occupies a  central position  in the  yeast $DB$  suggesting a
bow-tie causal organisation.

It is  worth to  note that, according  to its  definition, SCC
  components  may be  substantially modified  by the  addition  of new
  links or  by rewiring  them. In this context,  novel findings in
  the  gene regulatory  circuitry  of these  organisms  may alter  the
  current picture of these  networks. Now  an interesting  question arises.  From  an evolutionary perspective,
  mutations  in  regulatory  boxes  of  target genes,  may  alter  the
  structure  of  SCCs.  Then,  it  is reasonable  to  think  that  the
  regulatory  system   should  be  protected  against   this  kind  of
  perturbations. Interestingly, when comparing real networks with their
  null model counterparts  both $DB$ and SCCs are  markedly smaller than
  what is  expected  by  chance.  At  this point,  the  reduction  of  the
  dynamically relevant  core and the hierarchical  organisation of the
  smallest possible  size SCCs  (attending to dynamical  requisites of
  the  cell) would contribute  to reduce  the impact  of this  type of
  mutations.  In  addition,  another   layer  of  complexity  must  be
  considered to  play a relevant role  to minimise the  impact of this
  type of  mutations. This is  the distribution of the  activatory and
  inhibitory    weight     in    the    genetic     wiring. Interestingly,  it has seen  that an experimental  rewiring of
  the  \textit{E.  coli} gene  regulatory  network  does  not cause  a
  dramatic  impact on  organism surveillance  \citep{Isalan2008}.
 
Uncovering  the  genetic circuitry  of  these  organisms  is still  an
ongoing  task,   and  our  results   can  only  provide   a  tentative
approximation to the final picture.   However, due to the good quality
of data  for \textit{E.  coli}  network, only slight  modifications in
the  feed-forward architecture  are  expected.  In  the  case of  $DB$
yeast, its cyclic character is  a distinctive trait. Notice that the lack
of knowledge for gene connections would not alter the cyclic character
of $DB$. In addition, only a tree-like structure (similar to bacteria)
could be suggested if  further evidences demonstrate that current data
is enriched with  a high number of false  connections causing loops in
the network.  This  is a technical limitation in  yeast due to the fact that TF-gene
connections are defined according to  a statistic value [see yeast GRN
  construction in  Methods]. In our case,  very restrictive conditions
were used for yeast network construction in order to reduce the impact
of  possible  false   connections.   This  provides  strong  arguments
supporting that different strategies  for GRN organisation are present
in these prokaryotic and eukaryotic studied organisms.

Attending to the biology of bacteria and yeast, common patterns can be
identified. Both share an unicellular organisation, simple life cycles
(only sporulation  and replication states for yeast  and bacillus) and
similar  nutritional requirements (all  of them  are chemoheterotrophs
absorbing  molecules  from  the  medium).   However,  a  more  complex
behaviour appears associated to the  cell cycle control in yeast. As a
consequence,  these  relatively  simple  organisms may  operate  under
comparable  external inputs and  a similar  GRN organisation  would be
expected.   This is  consistent  with the  different output  behaviour
observed in  yeast and bacteria. It  is reasonable to  think that this
difference should originate by the different wiring displayed in these
GRNs.

Yeast and  bacteria drastically  differ in their  genomic organisation
and cellular  compartmentalisation, the two  most distinctive features
distinguishing  eukaryotes from  prokaryotes.  It  is  widely accepted
that spatial  compartmentalisation and intronic  organisation confer a
more              diverse             regulatory             behaviour
\citep{Lewis2003,Monk2003a,Swinburne2008,Swinburne2008a}  and  genomic
plasticity. GRN structures could be thus understandable as alternative
evolutionary solutions.  More complex  wiring in yeast, in contrast to
bacteria,  would  be the  basis  for  this differential  computational
behaviour.

\section{Methods}

\subsection{Construction of GRNs}

\subsubsection{E. coli network definition and construction}
GRN for E. coli is an overlapping of two files obtained from RegulonDB
6.0 \citep{Gama-Castro2008}: NetWorkSet.txt,  containing TFs and their
target  genes, and SigmaNetWorkSet.txt,  containing the  Sigma factors
and the genes  promoted by them. Both files  contain information about
the  relations, as well  as the  activator/repressor behaviour,  of TF
(and Sigma factors) over the target genes.  Biological information was
obtained  from  EcoCyc  database  \citep{Keseler2005}. In  this   work   we  exclude   elements
contributing to a TF modification such as phosphorylation or ligand-TF
binding.  Graph  pictures  were  performed  using  Cytoscape  software
(http://www.cytoscape.org/).

\subsubsection{{\em B. subtilis} network definition and construction}

GRN  for  {\em  B.   subtilis}  was obtained  by  the  combination  of
\textit{gene},  \textit{tfac}  and  \textit{sigma}  field  information
compiled   in  DBTBS   (release  5)   \citep{Sierro2008}.   Biological
information was obtained  from DBTBS, SubtiList \citep{Moszer2002} and
Uniprot \citep{Apweiler2004} databases.  Graph pictures were performed
using Cytoscape software (http://www.cytoscape.org/).

\subsubsection{{\em S. cerevisiae} network definition and construction}
Yeast  GRN was  obtained  from the  compilation  of different  sources
performed  by \citep{Balaji2006a}.   Self-interactions  were not initially
included in  that work  and they  were directly  provided  by the
authors.    Data  corresponds   with   highly  confident   experiments
($P=0.001$  and three positive  replicas). Biological  information was
obtained  from Saccharomyces Genome  database SGD  \citep{Hong2008} and
Uniprot.  Graph  pictures  were  performed  using  Cytoscape  software
(http://www.cytoscape.org/). 

\subsection{Null model construction and analysis}
Null model networks were obtained by a randomisation process of the real GRN consisting of twice of the number  of links rewiring events.
  A rewiring  event is realised by an end node interchange in
  two randomly selected pairs. Then, the  arrows of  the new
  pairs  are arbitrarily  inverted.  Once   obtained  the  new
  connections, the algorithm verifies that they were not previously in
  the   network. If the rewired connection matches with existent links, the solution is rejected and the algorithm proceeds to a selection of two new links.   Number   of  autoloops   is  kept   constant  in this randomisation process.   This   randomisation   permits   to   destroy   local
  correlations  keeping   the  number   of  nodes, links   and  degree
  distribution.  A  thousand of randomised  networks were constructed
  for each of the three GRNs. In addition, we also obtained null models
  for the set of nodes  corresponding with TF activity, i.e, the nodes
  with  $k_{out}>0$ (1000 randomised  networks for  each of  the three
  GRNs). $DB$ size,
  SCC size  components and  number of root  nodes were  calculated per
  null  model network. Frequencies  were normalised  by the  number of
  replicas (1,000 for each of the six conditions, i.e, one GRN and the
  respective TF subset for each of three studied organisms).

\subsection{Statistical analysis of biological functions}
Statistics  for the estimation  of the overabundance  of biological
  functions were estimated for the $DB$ set, root nodes and the dynamical
  module     sets     for      the     three     different     studied
  organisms.  Hypergeometric test  with Benjamini  and  Hochberg false
  discovery rate correction,  and selected significance level of
  0.05 was  applied using the  gene ontology biological  functions. In
  the case of \textit{B subtilis}, gene ontology annotation by parsing
  information  containing \textit{B.  subtilis} taxon  ID.  The source
  file,  $gene\_ association.goa\_ uniprot$  (submission date  4/25/2009),
  was                           obtained                          from
  $http://www.geneontology.org/GO.current.annotations.shtml$. Specific
  \textit{E. coli} GO annotation  was directly obtained from the same
  website.  The  analyses were  performed  using  Bingo 2.3  Cytoscape
  plugging  \citep{Maere2005}. Notice that, gene annotation for yeast is included by default in this package. Detailed information of statistical results is provided in SI2 supplementary
  file.

In  addition,  biological functions  of  the  $DB$ were  manually
  collected from respective organism's databases (EcoCyc and RegulonDB
  for \textit{E.coli}, DBTBS and Subtilist for \textit{B. subtilis} and
  SDG for \textit{S. cerevisiae}.

\subsubsection{Software implementation}
Dynamical  backbone  algorithm  and  dynamical module  detection  were
implemented in  a software package using Perl  language and unix/linux
commands.  This  package  is  available  for  linux/unix  platforms  as
supplementary material. Graph  pictures were performed using Cytoscape
software (http://www.cytoscape.org/).

\section{Mathematical appendix}

\subsection{Causality, dynamics and topology}

A directed  graph (or  digraph), ${\cal G}\equiv\{V_{\cal  G}, E_{\cal
  G}\}$,   is   constituted   by   a   set   of   vertices,   $V_{\cal
  G}\equiv\{v_1,...v_N\}$, and the set of edges linking them, $E_{\cal
  G}\equiv\{e_1,..., e_L\}$.  Formally, an edge from $v_i$ to $v_j$ is
described by an ordered pair $e_k=\langle v_i,v_j\rangle$, depicted by
an  arrow  in the  picture  of  the  graph $v_i\rightarrow  v_j$.   An
alternating  sequence of  vertices  $v_1,..., v_n\in  V_{\cal G}$  and
edges $e_1,..., e_{n-1}\in E_{\cal  G}$, defines a {\em directed walk}
in  a  digraph  ${\cal  G}$   \citep{Gross1998}  if  there  is  a  set
$e_1,...e_{n-1} \in E_{\cal G}$ such that, for all $i<n$, $e_i=\langle
v_i, v_{i+1}\rangle$.  We  denote a directed walk (if  it exists) from
$v_1$ to $v_n$ as $\pi(v_1,  v_n)$. We can also define a \textit{walk}
between  $v_1$ and  $v_n$,  i.e.,  a sequence  of  vertices and  edges
connecting  $v_1$ and  $v_n$  no  matter the  directed  nature of  the
graph.

The number of outgoing links of  a given vertex $v_i\in V_{\cal G}$ is known as {\em out-degree}
(denoted by  $k_{out}(v_i)$) whereas  the number of  incoming edges  is the
{\em  in-degree}  ($k_{in}(v_i)$). Once $k_{in}$ and $k_{out}$ have been defined, we can obtain the {\em average connectivity} or {\em average degree}, to be noted $\langle k \rangle$:
\begin{eqnarray}    
\langle k\rangle&=&\frac{1}{|V_{\cal G}|}\sum_{v_i \in V_{\cal G}}k_{in}(v_i)+k_{out}(v_i)\nonumber\\
&=&\frac{2|E_{\cal G}|}{V_{\cal G}}.
\end{eqnarray}

Furthermore,  given a  vertex  $v_i\in
V_{\cal  G}$, it is  interesting to  define the  set of  vertices {\em
  affecting} it, to be noted $\Gamma_i$:
\begin{equation}
\Gamma_i=\{v_k\in V_{\cal G}:\langle v_k,v_i\rangle \in E_{\cal G}\}.
\end{equation}

Attending  to their reachability,  a directed  graph can  display three
types of components, namely:
\begin{enumerate}
\item
The strongly connected components (SCC), which are the subgraphs for which
every       two       vertices       are      mutually       reachable
\citep{Gross1998,Dorogovtsev2003}.
\item
The  incoming components  (IC), composed  by the  set  of feed-forward
pathways starting from vertices without in-degree and ending in SCC.
\item
The outgoing  components (OC) where pathways starting from SCC
end in vertices without out-degree (see figure \ref{SCC}).
\end{enumerate}

\subsubsection{Threshold network model}

In  this subsection,  we  briefly define  the  toy model  used in  the
example provided  in figure (\ref{diagrama}). 

In a dynamical setting,  the state ${\bf \sigma}(t)=(\sigma_1(t), ...,
\sigma_N(t))$ of a system ${\bf  \sigma}$ formed by $N$ elements would
be updated under some class  of dynamical process.  An example of such
dynamics is a threshold-like equation,  in which the state of a vertex
$v_i\in  V_{\cal G}$  at time  $t+1$  is updated  by the  sate of  the
vertices $v_k\in \Gamma_i$, namely
\begin{equation}
\sigma_i(t+1)  = \Theta \left  ( \sum_{j=1}^N  \omega_{ij} \sigma_j(t) - \theta_i
\right )\nonumber
\end{equation}
where $\Theta(x)=1$ if $x>0$ and  zero otherwise. Here $\theta_i$ is a
threshold  and the  weights $\omega_{ij}  \in \{-1,0,+1\}$  define the
type of  interaction between  genes. If the  state of each  element is
Boolean, i.   e. $\sigma_i \in  \{0,1\}$ the previous  model provides,
for  a given  initial  state,  a closed  description  of the  system's
behaviour.  Here  the matrix $W=(\omega_{ij})$  captures the structure
and nature of  causal links.

The small threshold network depicted in figure (\ref{diagrama}) starts
from   an   initial   state  where   $\sigma_4(0)=\sigma_5(0)=1$   and
$\sigma_i(0)=0;\;(i\neq 4,5)$ for other  units, a cyclic attractor (of
period  $12$) is  obtained. Here  only  two elements  have a  non-zero
threshold,  namely $\theta_1=\theta_4=1/2$.   Arrows  and end  circles
indicate $\omega=+1$  and $\omega=-1$ links,  respectively. The global
state $M(t)$ is given by $M(t)=\sum_{j=1}^N \sigma_j(t)$ (thick lines,
b).

\subsection{Dynamical backbone pruning}

Consider  the  pruning function  $\Phi:{\cal  G}\rightarrow B$,  where
$B\subseteq {\cal G}$. This function  takes a directed graph as input,
being its output the graph without all the vertices having $k_{out}=0$
(and the links pointing to them).  Accordingly,
\begin{equation}
\Phi({\cal G})=B_1\{V_{B_1},  V_{E_1}\},
\end{equation}
where 
\begin{equation}
\left\{
\begin{array}{ll}
V_{B_1}\equiv\{v_k
\in  V_{\cal  G}  :  k_{out}(v_k)>0\}\nonumber\\
E_{B_1}\equiv\{  \langle
v_k,v_i \rangle  \in E_{\cal G}:  v_k, v_i \in V_{B_1}\}\nonumber.
\end{array}
\right.
\end{equation}
Thus, the
computation is an iterative operation:
\begin{eqnarray}
\Phi({\cal G})&=&B_1,\nonumber\\
B_2&=&\Phi(B_1)\nonumber\\
&...&\nonumber\\
B_n&=&\Phi(B_{n-1})\nonumber
\end{eqnarray}
The resulting graph  at the $i$-th iteration is  denoted by $B_i\equiv
B_i(V_{B_i}, E_{B_i})$ and the computation ends when no further vertex
elimination occurs, i.e.  $B_n=B_{n-1}$.   If, for some $i\leqslant n$
a vertex is eliminated and it has no connections with a node belonging
to  $B_{k\geqslant i}$  we  let  this vertex  {\em  alive}.  At  every
iteration, this collection of {\em  single root} vertices define a set
$r_i$ and, from  these sets, we build the set $R_i$  of all the single
root vertices found until the $i$-th step:
\begin{equation}
R_i=\bigcup_{k\leqslant i}r_k \label{R definition}
\end{equation}

We have now all the ingredients  to provide a formal definition of the
dynamical backbone of a directed graph ${\cal G}$, $DB({\cal G})$. Let
us  assume that,  when performing  recursively the  operation $\Phi$
over   a  directed  graph,   we  reached   the  stable   state,  i.e.,
$B_n=B_{n-1}$.  The  Dynamical Backbone is  a subgraph of  ${\cal G}$,
$DB\subseteq {\cal G}$ defined as:
\begin{equation}
DB({\cal G})\equiv DB(V_{DB({\cal G})}, E_{B_n})=DB(V_{B_n}\cup R_n, E_{B_n})
\end{equation}

Once  $DB({\cal G})$  is defined,  we can  also be  interested  in the
fraction of the net that exclusively displays feed-forward structures,
to be noted $DB'({\cal G})$,  which does not coincide exactly with the
graph  $\widehat{DB}({\cal G})$,  obtained  from the  set of  vertices
$v_i\in  V_{\cal G}\setminus  V_{DB(\cal  G)}$ and  the  set of  edges
$\{e_i=\langle  v_k,  v_j\rangle\in E_{\cal  G}:  v_k, v_j\in  V_{\cal
  G}\setminus V_{DB(\cal G)}\}$.  To  properly identify it, we need to
define the subgraph $L_n$ as  the set of connections linking $DB({\cal
  G})$ to  $\widehat{DB}({\cal G})$ and  the vertices linked  to them.
Note that this subgraph may  display many components. Its main feature
is that the links end in  vertices outside the $DB({\cal G})$ but they
come from vertices belonging to $DB({\cal G})$.  We obtain the maximal 
feed-forward subgraph $DB'({\cal G)}$ from
\begin{equation}
DB'({\cal G})=\widehat{DB}({\cal  G})\cup L_n. 
\end{equation}

\subsection{Dynamical Modules and Hierarchy}

Let  us  assume that  we  are in  the  $k$-th  connected component  of
$DB({\cal G})$.   The $i$-th dynamical module of  the $k$-th connected
component of the  $DB({\cal G})$, $DM^k_i$, is a  set of vertices (and
the directed edges among them) that constitutes an irreducible unit of
causal relations. As we said above, the existence of cycles inside the
$DB({\cal  G})$  is  responsible  for  the possible  non  trivial  net
dynamics. Thus, if the $k$-th component of the $DB({\cal G})$ is not a
single  root node, the  concept of  Dynamical Module  is topologically
equivalent  to the  SCC. As  there can  be a  more than  one dynamical
module in a directed graph,  we can define $\Delta_k({\cal G})$ as the
set  of dynamical  modules of  the $k$-th  component of  the $DB({\cal
  G})$.  Interestingly,  when the dynamical module is  contracted to a
single  vertex,  the resulting  graph  is  a  directed, acyclic  graph
(figure  \ref{DBB  definition}c).    Such  an  operation  is  commonly
referred  in the  literature  as  the {\em  condensation}  of a  graph
\citep{Gross1998}.   Accordingly,  we can  construct  a new  condensed
graph,
\begin{equation}
H_k({\cal   G})\equiv   H_k(V_{H_k}, E_{H_k}),
\end{equation}
where $V_{H_k}=\Delta_k({\cal  G})$ and $E_{H_k}$ is the  set of links
connecting  the  different  dynamical  modules.  In  other  words,  we
collapse the elements of every dynamical module into a single node and
we let the links connecting  different modules of the component of the
$DB({\cal G})$  we are working in.  Notice that, as  a consequence, we
cannot state that
\begin{equation}
H_k({\cal  G})\nsubseteq   {\cal   G}.
\end{equation}
Interestingly,  when we consider  these $DM$'s  as single  vertices of
$H_k({\cal G})$, we  obtain a feed-forward graph. It  is precisely the
feed-forward organisation what enables  us to define an order relation
among the elements  of $DB({\cal G})$.  Such an  order relation can be
interpreted as the dynamical hierarchy of the network's dynamical core
and it is defined among the different modules of $DB_k$.  If we define
$\Pi(H_k({\cal G}))=\{\pi^k_1,..,\pi^k_m\}$ as the set of all existing
directed walks  over all nodes of  $H_k({\cal G})$, we  can define the
order relation "$>$" as:
\begin{equation}
(DM^k_i>DM^k_j)\leftrightarrow    (\exists    \pi(DM^k_i,   DM^k_j)\in \Pi(H_k) )
\label{order}
\end{equation}
Such an order  relation is called the {\em  Transitive closure} of the
graph  $H_k({\cal  G})$ \citep{Gross1998}.  The  above order  relation
provides our definition of causal hierarchy.

\section{Supplementary files}
Supplemental information (SI). An excel file containing the biological
function of the $DB$ genes for the studied organisms obtaining for a manual curation process.

Supplementary information 2 (SI2). An excel file containing the statistical functional analysis for $DB$, the set of root nodes and dynamical modules for the three studied organisms. GRNs and respective $DBs$  are provided  as supplementary  material in  Cytoscape format.

Software package for $DB$ and $DM$ identification is provided as supplementary material (SI3).

Supplemetary files are provided upon request.

\section*{Acknowledgements}
We thank  the members of the Complex Systems Lab for  their fruitful suggestions.
This  work  was  funded   by  the  6th  Framework  project  ComplexDis
NEST-043241 (CRC),  James McDonnell Foundation (BCM) and  the Santa Fe
Institute (RVS).

\end{document}